\documentclass{ws-ijmpa}
\usepackage[super,compress]{cite}
\usepackage{graphicx}
\usepackage{subfigure}
\begin{document}
\markboth{M. Catillo \& L. Ya. Glozman}{Distribution law of the Dirac eigenmodes in QCD}

%
\catchline{}{}{}{}{}
%

\title{Distribution law of the Dirac eigenmodes in QCD}

\author{Marco Catillo}

\address{Institut f\"{u}r Physik, Universit\"{a}t Graz, Universit\"{a}tsplatz 5\\
Graz, 8010, Austria\\
marco.catillo@uni-graz.at}

\author{Leonid Ya. Glozman}

\address{Institut f\"{u}r Physik, Universit\"{a}t Graz, Universit\"{a}tsplatz 5\\
Graz, 8010, Austria\\
leonid.glozman@uni-graz.at}

\maketitle

\begin{history}
\received{Day Month Year}
\revised{Day Month Year}
\end{history}

\begin{abstract}
The near-zero modes of the Dirac operator are connected to
spontaneous breaking of chiral symmetry  in QCD (\emph{SBCS}) via the
Banks-Casher relation. At the same time the distribution of
the near-zero modes is well described by the Random Matrix Theory
(\emph{RMT}) with the Gaussian Unitary Ensemble (\emph{GUE}). Then it has become
a standard lore that a randomness, as observed through distributions
of the near-zero modes of the Dirac operator, is a consequence
of \emph{SBCS}. The higher-lying modes of the Dirac operator are not
affected by \emph{SBCS} and are sensitive to confinement physics and
related $SU(2)_{CS}$ and $SU(2N_F)$ symmetries. We study the
distribution of  the near-zero and higher-lying eigenmodes of
the overlap Dirac operator within $N_F=2$ dynamical simulations.
We find that both the distributions of the near-zero and higher-lying
modes are perfectly described by \emph{GUE} of \emph{RMT}. This means that randomness,
while consistent with \emph{SBCS}, is not a consequence of \emph{SBCS} and is
linked to the confining chromo-electric field. 

\keywords{Lattice QCD; chiral symmetry breaking; random matrix theory.}
\end{abstract}

\ccode{PACS numbers: 12.38.Gc, 11.30.Rd}


\section{\label{sec:intro}Introduction}

In QCD the $SU(N_F)_L \times SU(N_F)_R$ chiral symmetry of the 
Lagrangian is broken spontaneously by the quark condensate of the vacuum down to $SU(N_F)_V$. A density of the near-zero modes of the Euclidean Dirac operator is connected via the Banks-Casher relation \cite{bib:BanksCasher} with the quark condensate, that is an order parameter for spontaneous breaking of chiral symmetry (\emph{SBCS}). It is believed that in the low-energy domain around the chiral limit QCD can be described by an effective theory that involves the lowest excitations of the theory, the (pseudo) Goldstone bosons 

\begin{equation}
 U(x) = e^{i \pi(x)/F_\pi}.
 \label{eq:goldstone_bosons}
\end{equation}

The effective low-energy Lagrangian $\mathcal{L}_{eff}$ in an expansion in powers of derivatives of the field $U(x)$ and powers of quark masses is given as 

\begin{equation}
 \mathcal{L}_{eff} =\frac{F_{\pi}^2}{4}tr\left( \partial_{\mu} U(x)^{\dagger}\partial_{\mu} U(x)\right)- \Sigma Re\;\left[ e^{i\Theta/N_F}\right]tr\left( MU^{\dagger} (x)\right) + ...,
\label{eq:LagEff1}
\end{equation}

where $M$ is the mass matrix in a theory with $N_F$ degenerate flavors $M = m I$, $m$ is the mass of a single quark flavor and $I$ is a $N_F\times N_F$ identity. $\Theta$ is the vacuum angle and $\Sigma$ is the quark condensate. Then the effective low-energy  partition  for Euclidean QCD in a finite box with the volume $V$ is given as

\begin{equation}
\mathcal{Z}_{eff} = \int D[U] \exp \left\lbrace -\int_V d^4 x\; \mathcal{L}_{eff} \right\rbrace.
\label{eq:lepf}
\end{equation}

Thus, two constants $F_\pi$ and $\Sigma$ determine the leading term
of the effective Lagrangian and the higher-derivative terms generate corrections, involving powers of $1/L^2$, where $L$ is the linear size of the
box, and powers of $M$.

The interaction of Goldstone bosons is suppressed by their momenta. Consequently, one can parametrize the $U(x)$ field
as $U(x) = U_0U_1(x)$, where $U_0$ describes the zero-momentum modes $p=0$ and is space-independent and $U_1(x)$ describes the modes with $p \neq 0$. In the limit in which 

\begin{equation}
\frac{\Sigma\: m}{F_{\pi}^2} \ll \frac{1}{\sqrt{V}},
\label{eq:limit_chpt}
\end{equation}

the zero-momentum modes dominate and the effective partition function with the trivial theta-angle, $\Theta =0$, becomes

\begin{equation}
\mathcal{Z}_{eff} =  
A \int_{SU(N_F)} D[U_0] \exp \left\lbrace V\Sigma Re\; tr \left( MU_0^{\dagger} \right)\right\rbrace,
\label{eq:lepf2}
\end{equation}

where $A$ is a normalization constant.  Thus, QCD in the low-energy domain near the chiral limit with spontaneous breaking of chiral symmetry can be effectively described  within the $\epsilon$ -regime ($\Lambda_{QCD} L >> 1$, $m_\pi L << 1$) by a theory with the partition function above \cite{bib:SpectrumOfTheDiracOperatorAndRoleOfWindingNumberInQCD}.

At the same time it is known that these low-energy chiral properties of QCD can be described within a model that relies on randomly distributed weakly interacting instantons in the QCD vacuum \cite{bib:Shuryak,bib:Diakonov}. The randomness of the instanton distribution in Euclidean space-time is reflected in the distribution of the near-zero modes of the Euclidean Dirac operator, because within this model the exact quark zero modes, which are due to a zero-mode solution of the Dirac equation for a massless quark in the field of an isolated instanton, in an ensemble of the (weakly) interacting instantons become the near-zero modes.

Motivated by these observations it was suggested in Ref. \citen{bib:ShurVer} that the low-energy domain of QCD, related to spontaneous breaking of chiral symmetry, can be described by the chiral random matrix theory (\emph{chRMT}) with

\begin{equation}
\mathcal{Z}_{eff} = \int P(W)dW,
\label{eq:part_func_rmt}
\end{equation}

where $W$ is some random matrix such that the density probability distribution of $W$ for $N_F$ degenerate flavors is given by  

\begin{equation}
P(W)dW
 = \mathcal{N} \left( \det (D + m)\right)^{N_F} e^{-\frac{N\beta \Sigma^2}{4}tr(W^{\dagger}W)} dW.
\label{eq:PWdW}
\end{equation}

Here $dW$ is Haar measure and $D$ is Euclidean Dirac operator,

\begin{equation}
 D = \gamma_{\mu} (\partial_{\mu} + igA_{\mu}(x)).
\label{eq:dirac_op}
\end{equation} 

Choosing the chiral representation for the $\gamma$-matrices

\begin{equation}
\begin{split}
\gamma_k = 
\left(\begin{matrix}
0 & i\sigma_k\\
-i\sigma_k &0 \\
\end{matrix}\right),\; k = 1, 2, 3
\\
\gamma_4 = 
\left(\begin{matrix}
0 & 1 \\
1 & 0 \\
\end{matrix}\right),
\;
\gamma_5 = 
\left(\begin{matrix}
1 & 0 \\
0 & -1 \\
\end{matrix}\right),
\end{split}
\label{eq:gammamatrices}
\end{equation}

the Dirac operator, if the mass is set to zero, has the following structure:

\begin{equation}
D = 
\left(\begin{matrix}
0 & iW \\
iW^{\dagger} & 0\\
\end{matrix}\right).
\label{eq:DiracOperRMT}
\end{equation}

The Dirac operator in a finite volume (i.e. on the lattice) is a large $ N \times N$ matrix that is determined by the lattice size. If this matrix is random and recovers for $N \rightarrow \infty$ the Dirac operator in continuum,
then the low-energy properties of QCD, related to \emph{SBCS}, should be consistent with the \emph{chRMT}.

In Eq.~(\ref{eq:PWdW}) $\mathcal{N}$ is the normalization constant, $\Sigma$ is a parameter that it is not always related to the chiral condensate (not - if we are beyond the $\epsilon$ regime), and $\beta$ is the Dyson index which is determined by the symmetry properties of the matrix $W$. Different values of $\beta$ correspond to different matrix ensembles. If $\beta =1$ we have the \emph{chiral Gaussian Orthogonal Ensemble} (\emph{chGOE}), if $\beta =2$ the \emph{chiral Gaussian Unitary Ensemble} (\emph{chGUE}) and $\beta = 4$, the \emph{chiral Gaussian Symplectic Ensemble} (\emph{chGSE}). In \emph{QCD} $\beta=2$ as was shown in Ref. \citen{bib:Ver}. 

Subsequent lattice studies of distributions of the lowest-lying modes in QCD have confirmed that these distributions follow a universal behaviour imposed by the Wigner-Dyson Random Matrix Theory \cite{bib:Universality-RMT,bib:Farchioni-Lang-Splittorff,bib:Luscher-Giusti,bib:Fukaya-JLQCD,bib:Edwards1,bib:Edwards2}. As a result it has become an accepted paradigm that  randomness of the low-lying modes is a consequence of \emph{SBCS}.

The lowest-lying modes of the Dirac operator are strongly affected by \emph{SBCS}. At the same time the higher-lying modes are subject to confinement physics. This was recently established on the lattice via truncation of the lowest modes of the overlap Dirac operator from the quark propagators \cite{bib:Denissenya1,bib:Denissenya2,bib:Denissenya3,
bib:Denissenya4}. Hadrons (except for pion) survive this truncation and their mass remains large. Not only
$SU(2)_L \times SU(2)_R$ and $U(1)_A$ chiral symmetries get restored, but actually some higher symmetries emerge. These symmetries were established to be $SU(2)_{CS}$ (chiral-spin) and $SU(4)$ that contain chiral symmetries as subgroups and that are symmetries of confining chromo-electric interaction \cite{bib:G1,bib:G2}. 
 
\begin{figure}[htbp]
\centerline{\includegraphics[scale=0.75]{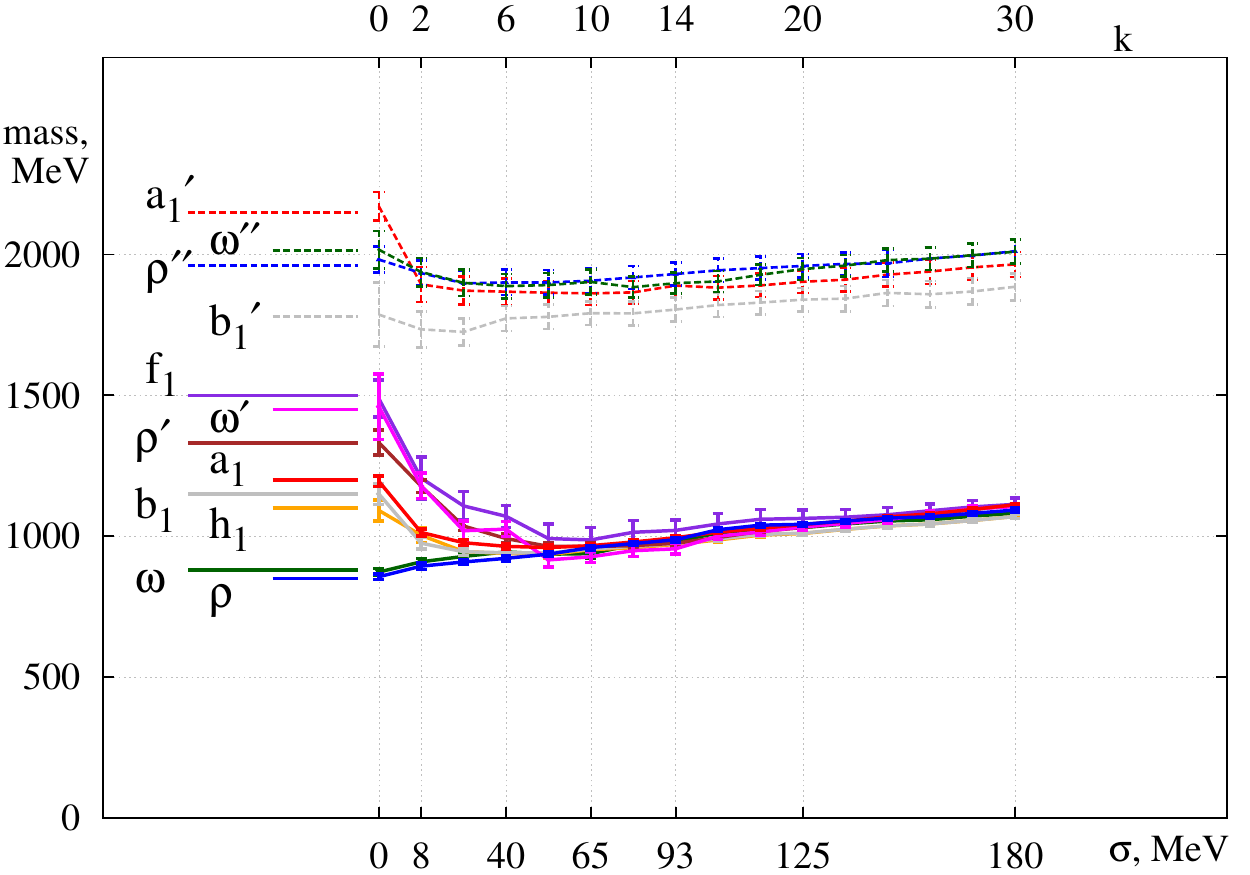}}
\caption{Mass evolution of $J = 1$ mesons on exclusion of the near-zero modes of the Dirac operator; $k$ is the number of truncated lowest-lying modes. The value $\sigma$ denotes the energy gap in the spectrum of the Dirac operator. Fig. from Ref. \citen{bib:Denissenya2}. \label{fig-Denissenya_plot}}
\end{figure}

The mass evolution of the $J=1$ mesons upon  truncation of $k$ lowest eigenmodes of the Dirac operator is shown on Fig. \ref{fig-Denissenya_plot} \cite{bib:Denissenya2}. It is obvious that information about $SU(2)_L \times SU(2)_R$ and $U(1)_A$ breakings is contained in lowest 10-20 modes (given the lattice size $L \sim 2$ fm). The higher-lying modes reflect a $SU(2)_{CS}$  and $SU(4)$ symmetric regime and are not sensitive to \emph{SBCS}.

Given success of \emph{chRMT} for the lowest-lying modes of the Dirac operator it is  natural to expect that the distribution law of the higher-lying modes should be different and should reflect confinement physics. This motivates our study of the distribution of the lowest-lying and higher-lying modes of the Dirac operator and their comparison.

\section{\label{sec:Lattice-Setup}Lattice Setup}

We compute 200 lowest eigenmodes of the overlap Dirac operator 
(see Refs. \citen{bib:Neuberger1,bib:Neuberger2})  

\begin{equation}
D_{ov} (m) = \left( \rho + \frac{m}{2} \right)  + \left( \rho - \frac{m}{2} \right) \gamma^5 sign \left[  H(-\rho) \right] ,
\label{eq-overlap}
\end{equation}

where $H(-\rho) = \gamma^5 D(-\rho)$ and $D(-\rho)$ is the Wilson-Dirac operator; $m = 0.015$ is 
the valence quarks mass and $\rho = 1.6$ is a simulation parameter. 
The overlap operator is $\gamma^5$-hermitian

\begin{equation}
D_{ov} (0)^{\dagger}= \gamma^5 D_{ov} (0) \gamma^5
\label{eq:dir_op_props}
\end{equation}

and satisfies the Ginsparg-Wilson  relation 

\begin{equation}
\lbrace \gamma^5,D_{ov} (0) \rbrace = \frac{1}{\rho} D_{ov}(0)\gamma^5 D_{ov}(0).
\label{eq-GW}
\end{equation}

The eigenvalues of the overlap Dirac operator lie on a circle with radius $R= \rho - \frac{m}{2}$, see Fig. 2, and come in pairs $(\lambda_{ov}(m),\lambda_{ov}^{*}(m))$. This is a consequence of Eq. (\ref{eq-GW}) and the $\gamma^5$-hermiticity. Hence the eigenvalues below the real axis bring the same informations as the eigenvalues above the real axis. For this reason we consider, for our analysis, only the eigenmodes with $Im(\lambda_{ov} (m)) \geq 0$. 

In order to recover the eigenvalue $\lambda$ of the massless Dirac operator in continuum theory we need to project 
our eigenvalues on the imaginary axis. There is no unique way to define projection. For this purpose we consider three different definitions. All of these definitions are illustrated on Fig. 2. We will study the sensitivity of our results on choice of projection definition. For reasonably low eigenvalues and not large quark masses we don't expect a large variation in so defined projections.

We use 100 gauge field configurations in the zero  global topological charge sector generated by JLQCD collaboration 
with $N_F = 2$ dynamical overlap fermions on a $L^3 \times L_t  = 16^3 \times 32$ lattice
with $\beta = 2.30$ and lattice spacing $a \sim 0.12fm$. The pion mass is $m_{\pi} = 289(2)MeV$, see Refs. \citen{bib:JLQCD1,bib:JLQCD2}. Precisely the same gauge configurations  have been used in truncation studies \cite{bib:Denissenya1,bib:Denissenya2,bib:Denissenya3,bib:Denissenya4}. 

The eigenvalues $\lambda_{ov}(m)$ of $D_{ov} (m)$ are obtained calculating, at first, the sign function $ sign [H]$. 
We use the Chebyshev polynomials to approximate $ sign [H]$ with an accuracy of $ \epsilon = 10^{-18} $, and then  compute $ 200 $ eigenvalues of $ D_{ov} (m) $.

We notice that with this lattice setup it is not a priori obvious that \emph{chRMT} should work, because we are beyond the $\epsilon$-regime, in our case $m_{\pi}L \simeq 3$. 

\begin{figure}[htbp]%
\centerline{
\subfigure[$ \lambda = \frac{Im\:\lambda_{ov}(m)}{1 - \frac{Re\:\lambda_{ov}(m)}{\rho + \frac{m}{2}}} $]{%
\label{fig-PsCenterGraph}%
\includegraphics[scale=0.85]{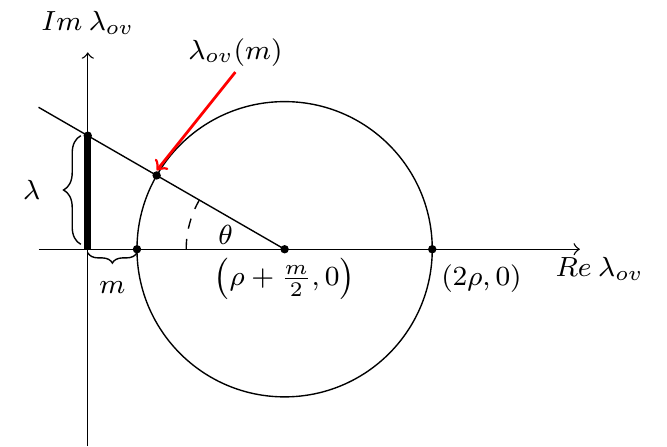}
}%
\subfigure[$ \lambda = \frac{Im\:\lambda_{ov}(m)}{1 - \frac{Re\:\lambda_{ov}(m)}{2\rho}}$]{%
\label{fig-Ps2rhoGraph}%
\includegraphics[scale=0.85]{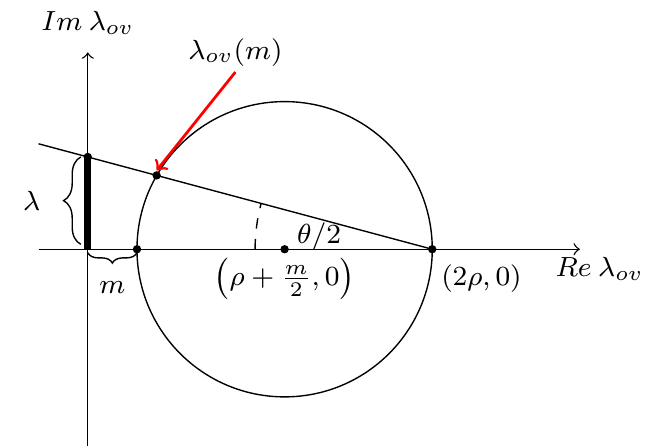}
}%
\\
\subfigure[$\lambda = \left( \rho - \frac{m}{2}\right)\theta $]{%
\label{fig-PsThetaGraph}%
\includegraphics[scale=0.85]{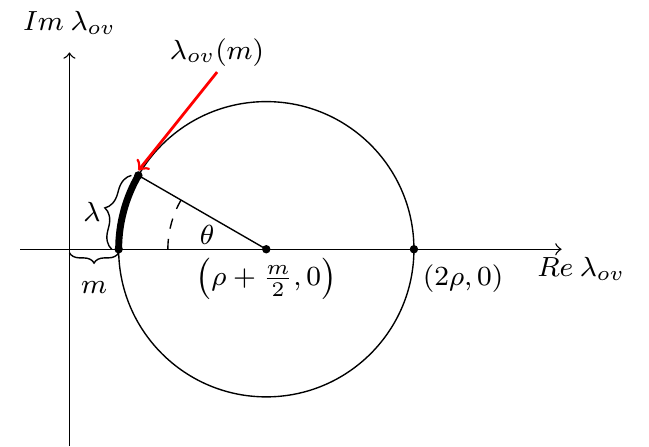}
}%
}
\caption{Different definitions of $\lambda$  using the eigenvalues of the overlap Dirac operator, $\lambda_{ov} (m)$.
The angle $\theta$ is defined as $\theta = arctg\left( \frac{Im\: \lambda_{ov}(m)}{\left( \rho + \frac{m}{2}\right) - Re\: \lambda_{ov}(m)}\right) $. \label{fig-ProjDefinition}}
\end{figure}

\section{\label{sec:Lowest-eigenvalues}Lowest Eigenvalues}

As we have seen we are out from $\epsilon$-regime and the full agreement with \emph{chRMT} is not a priori expected. 
Nevertheless we want to check whether \emph{chRMT} can still describe the lowest eigenvalues in our system. 

An important prediction of \emph{chRMT} is the distribution of the lowest eigenvalues in the limit when the four  dimensional volume $V\rightarrow \infty$ and the quantity $V\Sigma m_{\pi}$ is fixed, see Ref. \citen{bib:Dam}. We  order the projected eigenvalues such that $\lambda_1 \leq \lambda_2 \leq ... \leq \lambda_N$, then we define the variables $\zeta_k  = V\Sigma \lambda_k$ and get the distribution $p_k (\zeta_k)$ of each $\zeta_k$ (Ref. \citen{bib:Dam}). 

In Table \ref{tab-EigRatio} we show the ratios $\langle \lambda_k \rangle/\langle \lambda_j \rangle$, for $1 \leq j < k \leq 4$, where $\langle \lambda_i \rangle$ is the average over all gauge configurations for the $i$th projected eigenvalue. Since we don't know the parameter $\Sigma$, we can use that $\langle \zeta_k \rangle = V\Sigma \langle\lambda_k \rangle$ and we can compare our ratios with the predictions of \emph{chRMT}. 

\begin{table}[ph]
\tbl{Ratio $\langle \lambda_k \rangle / \langle \lambda_j \rangle$ for $1\leq j \leq k \leq 4$ and the same values computed with the \emph{chRMT}. We denote with $\sigma$ the error. In this case we have used $\lambda$ defined as in Fig. \ref{fig-Ps2rhoGraph}.}
{\begin{tabular}{@{}cccc@{}} \toprule
  		$k/j $ & $\langle \lambda_k \rangle / \langle \lambda_j \rangle$ & $\sigma$  & \emph{chRMT} \\ \colrule 
  		2/1 & 2.72 &	 0.19 & 2.70 \\
  		3/1 & 4.35 &	 0.28 & 4.46 \\
  		3/2 & 1.60 & 0.06 & 1.65 \\
  		4/1 & 5.92 &	 0.38 & 6.22 \\
  		4/2 & 2.17 &	 0.08 & 2.30 \\
  		4/3 & 1.36 &	 0.03 & 1.40 \\ \botrule
  	\end{tabular}\label{tab-EigRatio}}
\end{table}

\begin{figure}[htbp]
\centerline{\includegraphics[scale=0.5]{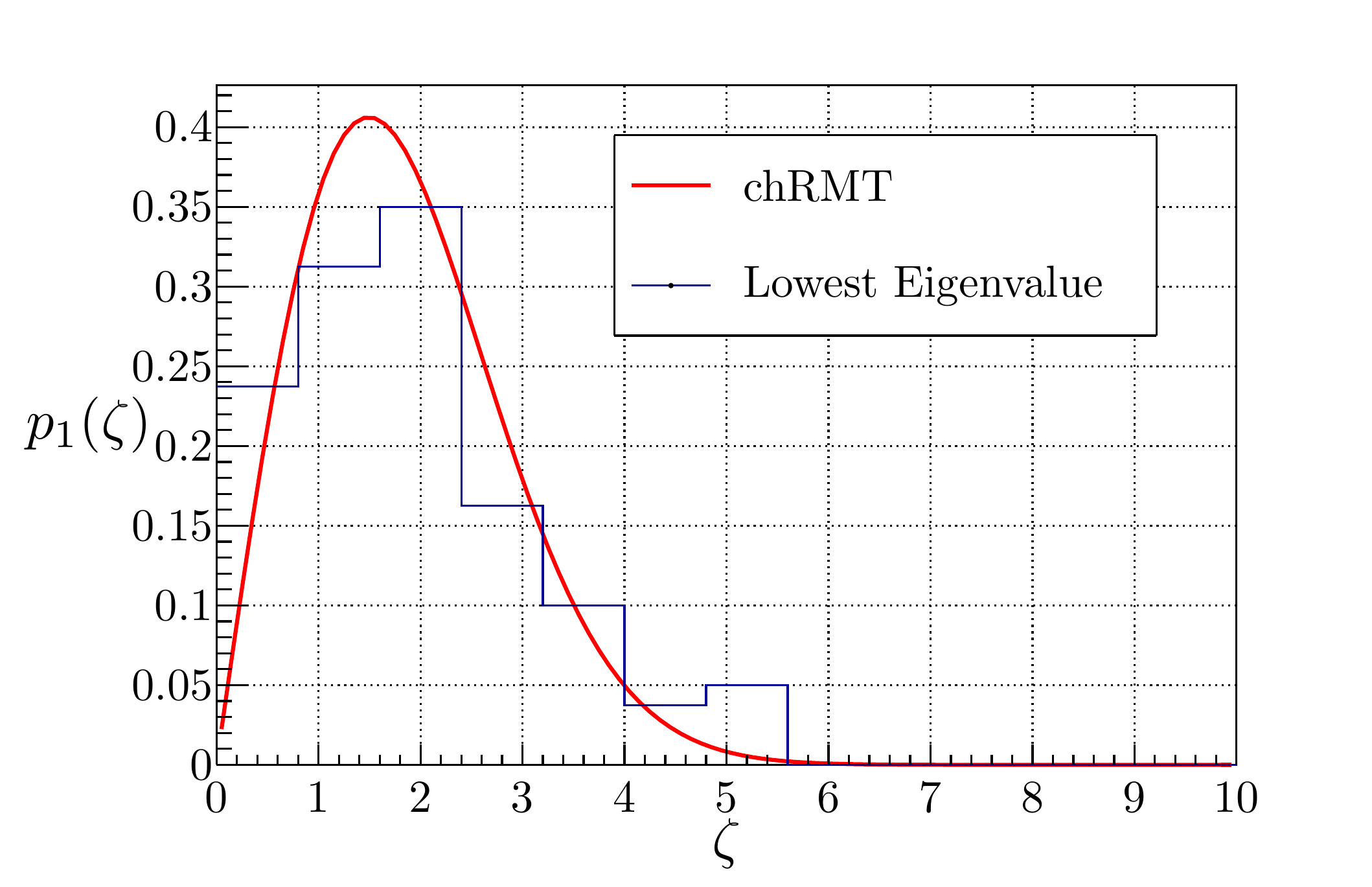}}
\caption{Distribution of the lowest eigenvalue. In this case $\zeta = V\Sigma \lambda$. \label{fig-Distr1stEig}}
\end{figure}

We  see that the ratios for the first 3 projected eigenvalues are in good agreement with the \emph{chRMT}.
The ratios involving the 4-th projected eigenvalues have a larger discrepancy. From the theoretical values of $\langle \zeta_k \rangle$ and the observed values of $\langle \lambda_k \rangle$ we can extract the parameter $\Sigma$. We find $\Sigma = (232.2 \pm 0.9 MeV)^3$. We use this parameter to compare the distribution $p_1(\zeta) = dN_1/d\zeta$ of the first lowest projected eigenvalue with the theoretical distribution given by \emph{chRMT}, as we report in Fig. \ref{fig-Distr1stEig}. $dN_1$ is the number of values assumed by the lowest projected eigenvalue of the Dirac operator, multiplied by $V\Sigma$, for different configurations in the interval $(\zeta,\zeta + d\zeta)$. We  conclude this section noting that, even though we are not in the $\epsilon$-regime, for very low eigenvalues  the predictions of \emph{chRMT} are in good agreement with data.

\section{\label{sec-NNS}Nearest Neighbor Spacing Distribution}

In this section we consider another important prediction of \emph{chRMT}. We first define the variable 

\begin{equation}
s_n = \xi_{n+1}-\xi_n,
\label{eq-s_n}
\end{equation}

where

\begin{equation}
\xi_n =  \xi (\lambda_n) = \int_0^{\lambda_n}R(\lambda)d\lambda
\label{eq-xi_n}
\end{equation}

\noindent
and $R(\lambda)$ is the probability to find an eigenvalue of the Dirac operator inside the interval $(\lambda,\lambda +d\lambda)$. $n$ indicates the number of the lowest projected eigenvalue, supposing we have ordered them in ascending order as described in the previous section. The distribution of the variable in Eq. (\ref{eq-s_n}) is called nearest neighbor spacing distribution (or \emph{NNS} distribution). 

In principle we don't have access to the theoretical distribution $R(\lambda)$ and the calculation of $\xi (\lambda_n)$ is not trivial. The procedure to map the set of variables $\lbrace \lambda_1, ...,\lambda_N \rbrace$ into the set $\lbrace \xi_1,...,\xi_N \rbrace$ is called \emph{unfolding} and it is described in Ref. \citen{bib:Guhr}. To unfold we introduce the following variable 

\begin{equation}
\begin{split}
\eta (\lambda_n) =& \int_{0}^{\lambda_n} \rho (\lambda)d\lambda = \frac{1}{N} \langle \sum_k \theta (\lambda_n - \lambda_k) \rangle = \\
=& \frac{1}{N}\frac{1}{M}\sum_{i=1}^M \sum_k \theta (\lambda_n - \lambda_k^i),
\end{split}
\label{eq-eta_n}
\end{equation}

\noindent
where $\rho (\lambda) =  \frac{1}{N} \langle \sum_k \delta (\lambda - \lambda_k) \rangle$ is the spectral density of the Dirac operator averaged over all gauge field configurations, $\lambda_k^i$ denotes the $k$th lowest projected eigenvalue of the Dirac operator computed using the $i$th gauge configuration and $M = 100$ is the number of gauge configurations. It is shown in Ref. \citen{bib:Guhr} that we can decompose $\eta (\lambda)$ into a global smooth part $\xi (\lambda)$ and a local fluctuating part $\eta_{fl}(\lambda)$:

\begin{equation}
\eta (\lambda) = \xi (\lambda) + \eta_{fl}(\lambda).
\label{eq-eta_xi}
\end{equation}

\begin{figure}[htbp]
\centering
\centerline{\includegraphics[scale=0.5]{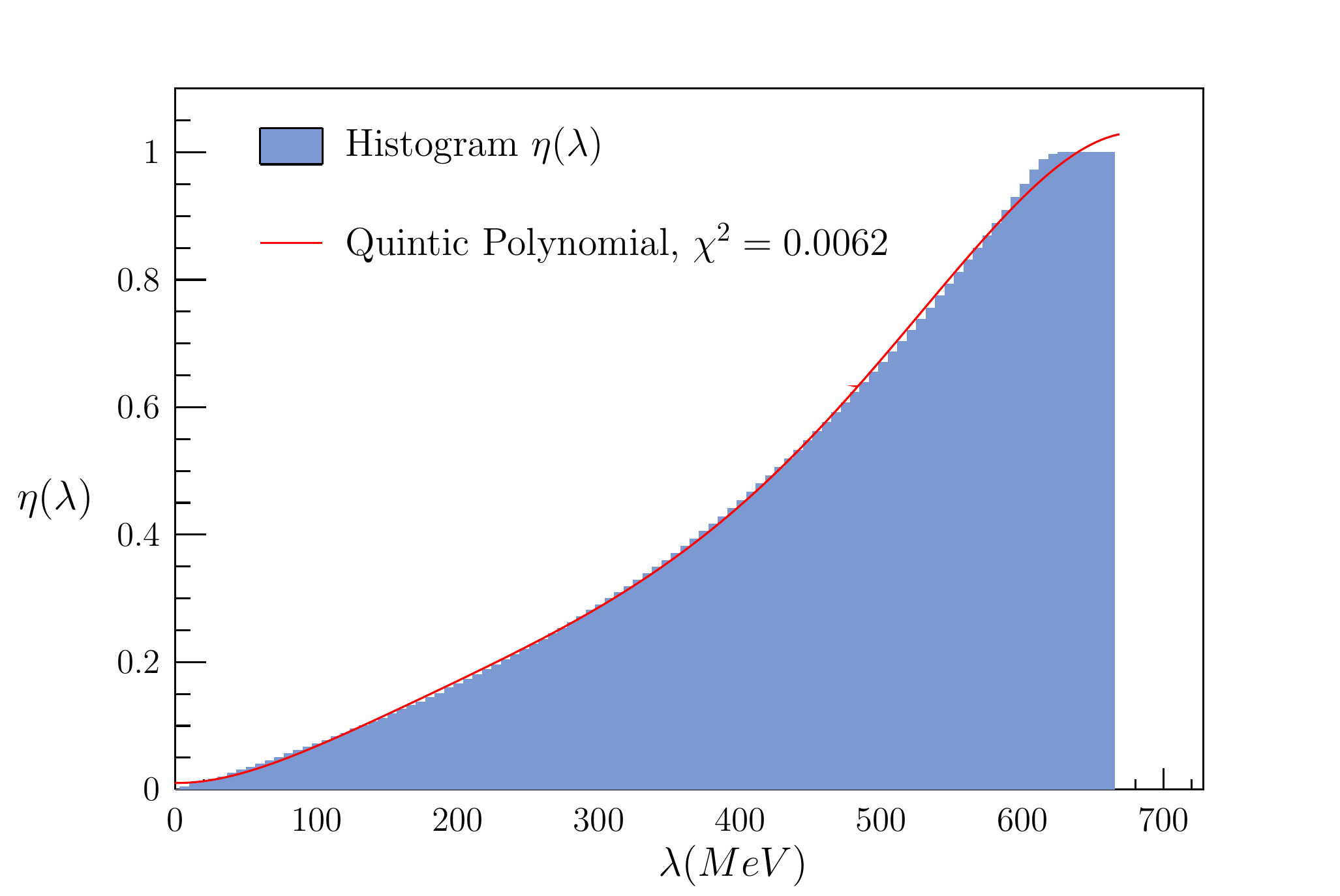}}
\caption{\emph{Unfolding procedure} normalized to the total number of the calculated eigenvalues of the Dirac operator. \label{fig-UnfProc}}
\end{figure}

\noindent
The smooth part can be obtained by a polynomial fit of $\eta (\lambda)$, as shown in Fig. \ref{fig-UnfProc}. 

For different values of the Dyson index $\beta$ we have different shapes of the \emph{NNS} distribution. 

We use the \emph{NNS} distribution $p(s) = dN_s/ds$, where $dN_s$ represents the number of the values $s_n$ inside the interval $(s,s+ds)$, to study the lowest and the higher eigenvalues of the overlap Dirac operator. 

The lowest eigenvalues contain the information about $SU(2)_L \otimes SU(2)_R \otimes U(1)_A$ breaking as is evident from Fig. \ref{fig-Denissenya_plot}. On the other hand  the higher-lying eigenvalues, with $k > 10-20$  are not sensitive to \emph{SBCS} and to breaking of $U(1)_A$, but reflect  physics of confinement and of   $SU(2)_{CS}$ and $SU(4)$ symmetries. 

\begin{figure}[htbp]
\centerline{
 \subfigure[Range eigenvalues: 1 - 10]{
 \label{fig-1-10}
 \includegraphics[scale=0.35]{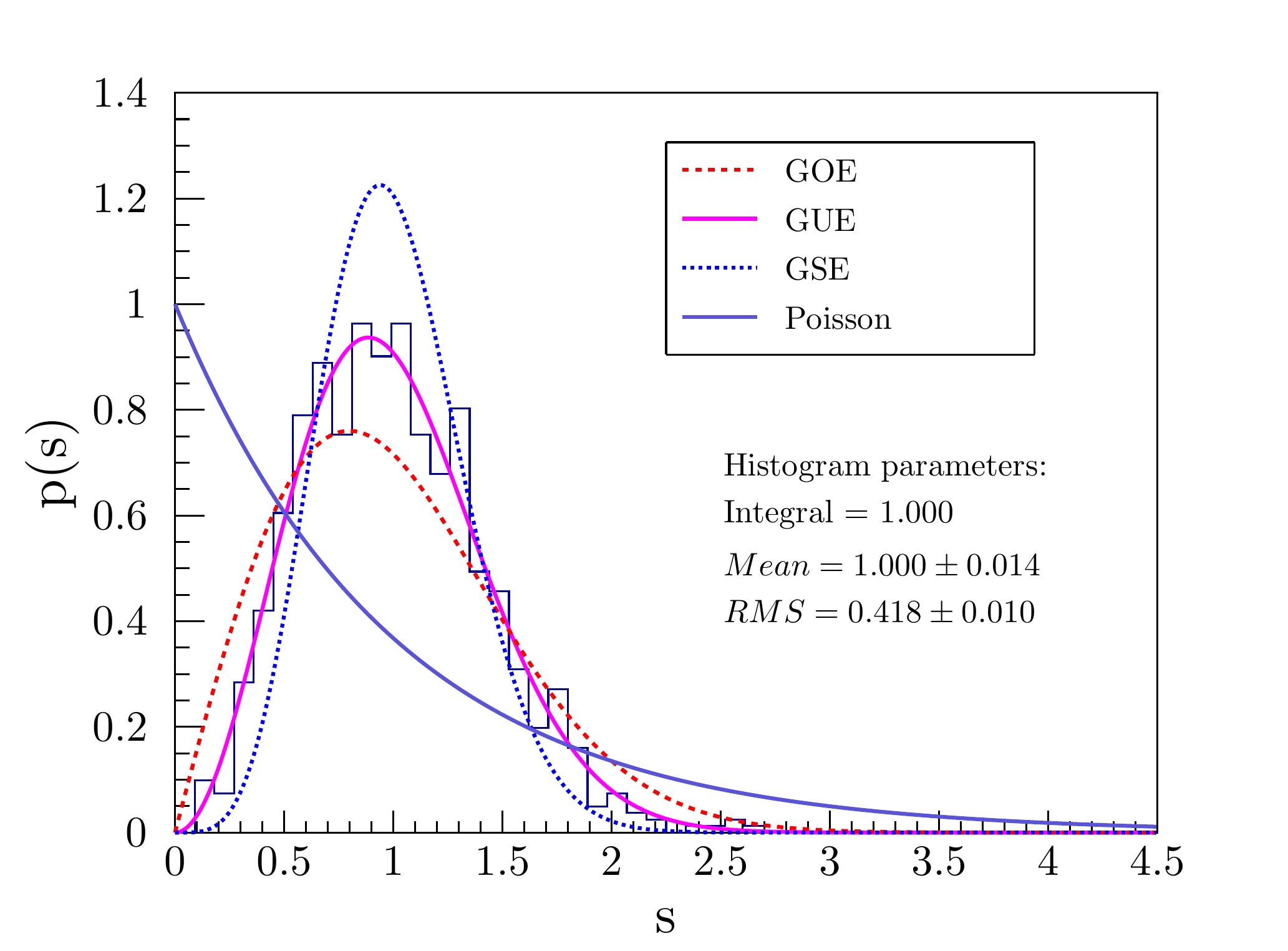}
 }\quad
\subfigure[Range eigenvalues: 81 - 100]{
  \label{fig-81-100}
  \includegraphics[scale=0.35]{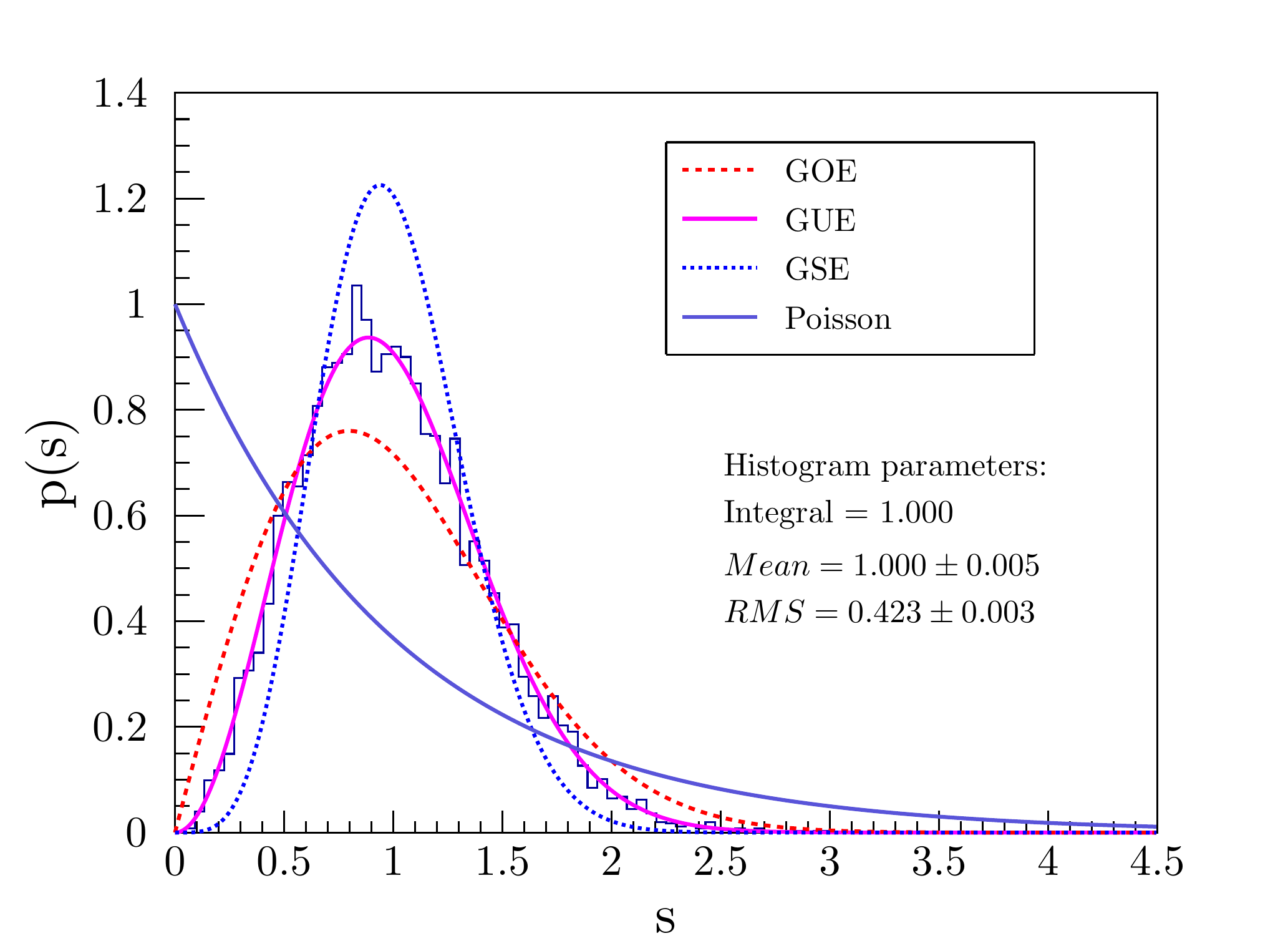}
 }}
 \caption{\emph{NNS} distribution for the first $10$ lowest projected eigenvalues of the overlap Dirac operator (a) and the higher eigenvalues (b). We have used $\lambda$ defined as in Fig. 1(b). \label{fig-NNS_distribution}}
\end{figure}

The \emph{NNS} distributions obtained with the lowest 10 eigenmodes and with the eigenmodes in the interval 81 - 100 are shown on Fig. \ref{fig-NNS_distribution}. We see that the distribution of the lowest 10 eigenmodes is perfectly described by the Gaussian Unitary Ensemble, in agreement with \emph{chRMT}. However, the same Wigner distribution is observed for the higher-lying eigenmodes, which is unexpected. This tells that the Wigner distribution is not a consequence of \emph{SBCS} in QCD and has a more general root.

\begin{figure}[htbp]
\centerline{
 \subfigure[$ \lambda = \frac{Im\:\lambda_{ov}(m)}{1 - \frac{Re\:\lambda_{ov}(m)}{\rho + \frac{m}{2}}} $]{
 \label{fig-PsCenter}
 \includegraphics[scale=0.35]{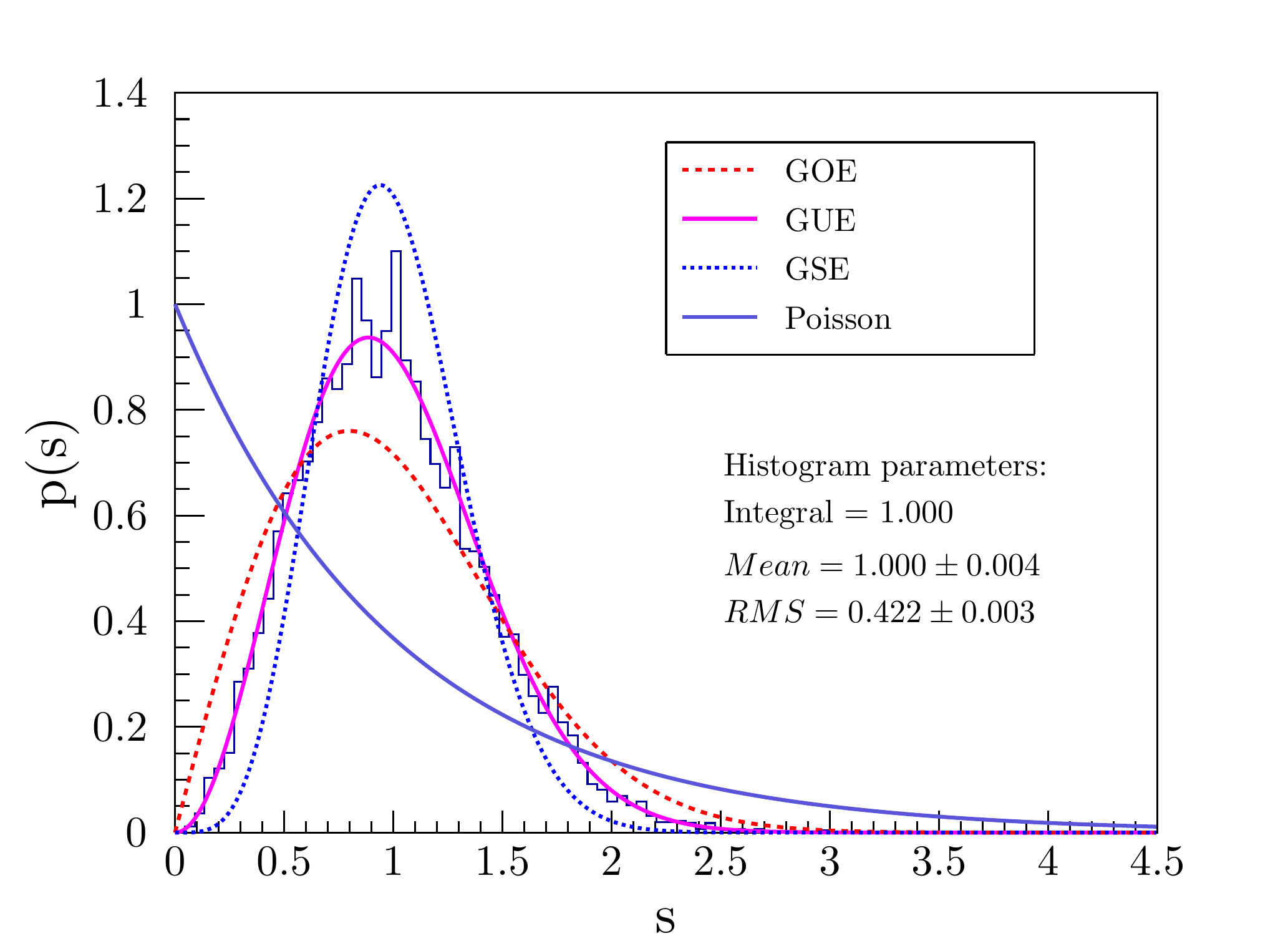}
 }
 \subfigure[$ \lambda = \frac{Im\:\lambda_{ov}(m)}{1 - \frac{Re\:\lambda_{ov}(m)}{2\rho}}$]{
 \label{fig-Ps2rho}
 \includegraphics[scale=0.35]{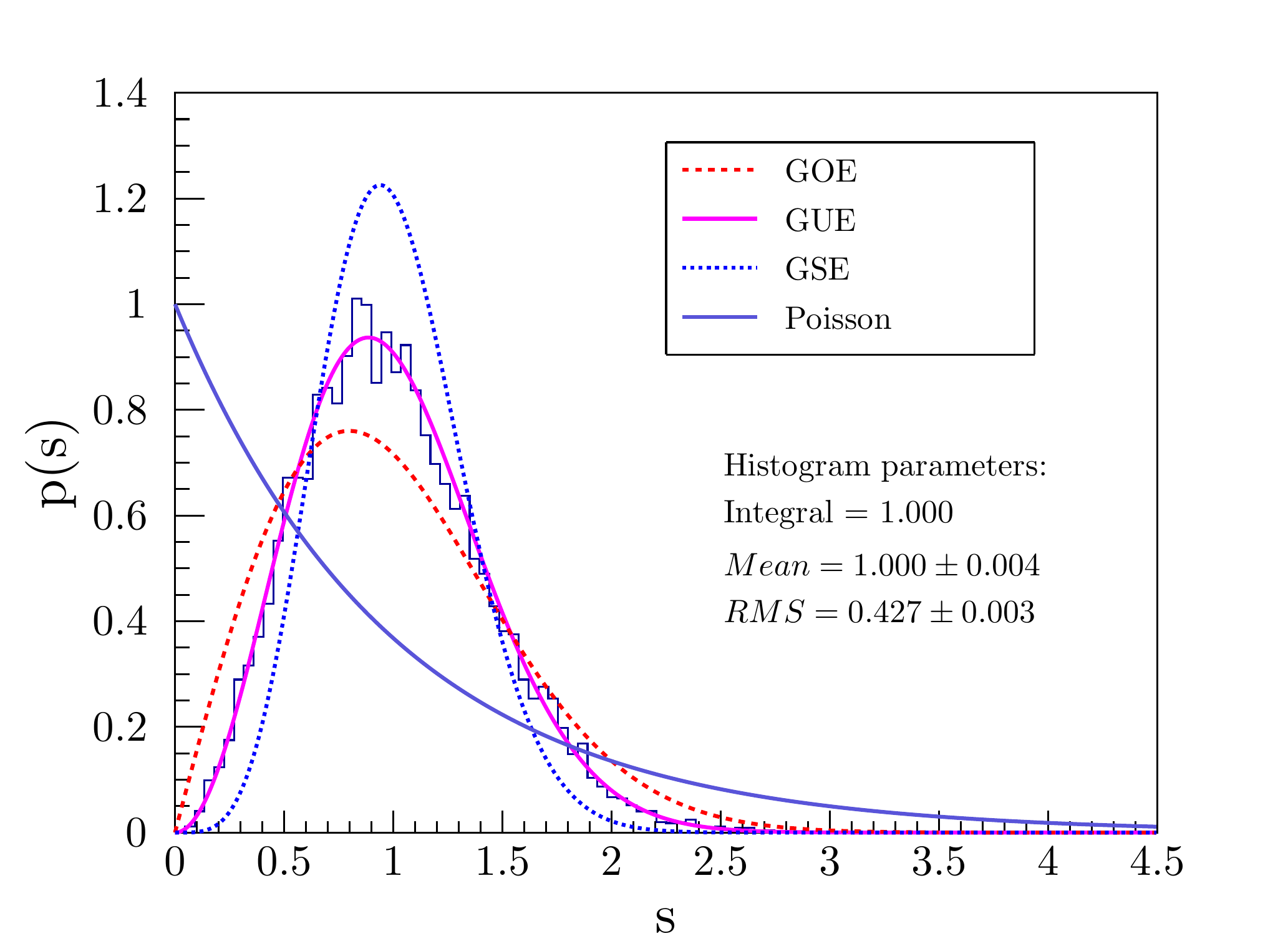}
 }}
 \centerline{
 \subfigure[$\lambda = \left( \rho - \frac{m}{2}\right)\theta $]{
 \label{fig-PsTheta}
 \includegraphics[scale=0.35]{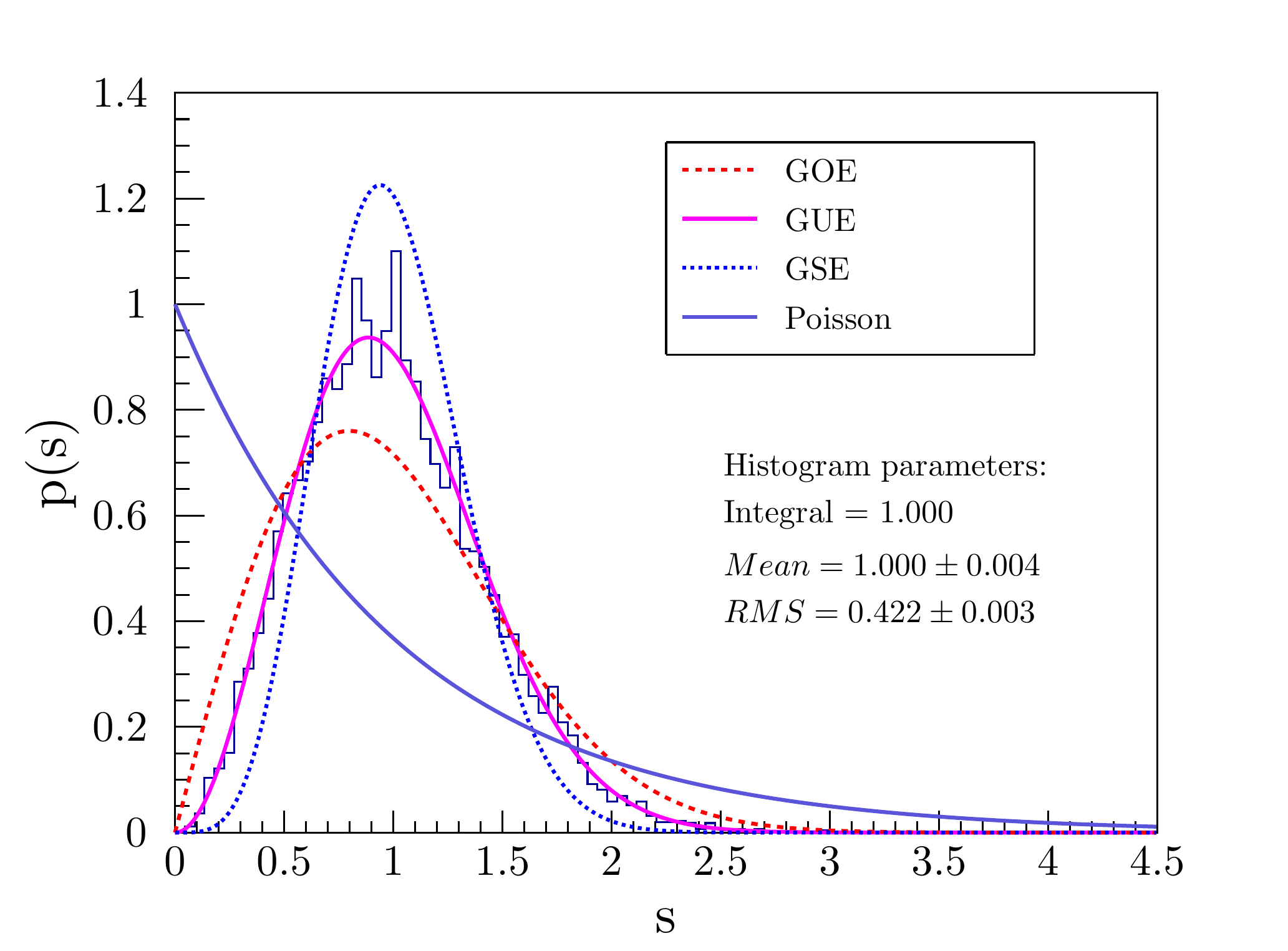}
 }}
  \caption{\emph{NNS} distribution for the lowest $100$ eigenvalues of the Dirac operator, for different definitions of $\lambda$. \label{fig-NNS_distribution_different_projections}}
 
\end{figure}

Finally, in Fig. \ref{fig-NNS_distribution_different_projections} we show \emph{NNS} distributions of the lowest 100 modes calculated with three different definitions of projected eigenvalue, compare with Fig. 1. It is clear that results for the distribution is not sensitive to definition of projected eigenvalue.

\section{\label{sec:conclusions}Discussion and Conclusions}

In the past the  distributions were studied typically for the low-lying modes, see e.g. Ref. \citen{bib:Universality-RMT,bib:Farchioni-Lang-Splittorff,bib:Luscher-Giusti,bib:Fukaya-JLQCD,bib:Edwards1,bib:Edwards2}. The observed Wigner distribution was linked to the \emph{SBCS} phenomenon. The \emph{NNS} distribution for all eigenmodes has been investigated on a small $6^3 \times 4$ lattice with the
staggered fermions in Ref. \citen{P}. The Wigner surmise has been noticed in this study (see also Ref. \citen{H}). On a small lattice it is difficult to distinguish the near-zero modes, responsible for \emph{SBCS} and the bulk modes, however.

We have used a reasonably large lattice with the chirally-invariant Dirac operator. More important, given our previous works \cite{bib:Denissenya1,bib:Denissenya2,bib:Denissenya3,bib:Denissenya4}, we have a control over which modes can be considered as the near-zero modes, that are related with \emph{SBCS}, and which are the bulk modes that are not affected by \emph{SBCS}. It is clear from the Fig. \ref{fig-Denissenya_plot} that on our lattice the physics of \emph{SBCS} is contained roughly in 10 lowest modes of the Dirac operator.The higher-lying modes are subject to confinement physics and related $SU(2)_{CS}$ and $SU(4)$ symmetries. The higher-lying modes do not carry information about \emph{SBCS}.

We have found that even beyond the $\epsilon$-regime \emph{RMT} describes well the lowest eigenvalues of our system in agreement with previous results. We have also found that the higher-lying modes, that are not sensitive to \emph{SBCS},  follow the same Wigner distribution as the near-zero modes. 

This observation means that the Wigner distribution seen  both for the near-zero and higher-lying modes, while consistent with spontaneous breaking of chiral symmetry, is not a consequence of spontaneous breaking of chiral symmetry in QCD but has some more general origin in QCD in confinement regime.

An interesting question is what part of the QCD dynamics is primarily linked to randomness. We can answer this question given the new $SU(2)_{CS}$ and $SU(4)$ symmetries and their connection to a specific part of the QCD dynamics \cite{bib:G1,bib:G2}. In particular, it is the chromo-electric part of the QCD dynamics that is a source of these symmetries. At the same time the chromo-magnetic interaction breaks both symmetries. This symmetry classification  distinguishes different parts of the QCD dynamics. The emergence of the $SU(2)_{CS}$ and $SU(4)$ symmetries upon truncation of the near-zero modes of the Dirac operator allows to claim that the effect of the chromo-magnetic interaction in QCD is located exclusively in the near-zero modes, while confining chromo-electric interaction is distributed among all modes of the Dirac operator. 

Obviously some  microscopic dynamics should be responsible for this. Given our observation that both the near-zero and the bulk modes are subject to randomness, we can conclude that some unknown random dynamics in QCD is linked to the confining chromo-electric field. This conclusion is reinforced by a recent study \citen{R} of high temperature QCD where the near-zero modes of the Dirac operator are suppressed and where the chiral symmetry is restored. There the same $SU(2)_{CS}$ and $SU(4)$ symmetries are observed and the results indicate that the notion of ``trivial" deconfinement (related to the Polyakov loop) has to be reconsidered.

\section*{Acknowledgments}

We are grateful to  C.B. Lang for numerous discussions and thank  J. Verbaarschot for careful reading of the paper. We also thank the JLQCD collaboration for supplying us with the overlap gauge configurations. This work is supported by the Austrian Science Fund FWF through grants  DK W1203-N16 and P26627-N27.


\end{document}